\documentclass[letter]{aa} 

\newcommand{\Msun}{\ensuremath{M_{\odot}}}
\newcommand{\Rsun}{\ensuremath{R_{\odot}}}
\newcommand{\Ni}{\ensuremath{\mathrm{^{56}Ni}}}

\newcommand{\ej}{\ensuremath{\mathrm{ej}}}
\newcommand{\Co}{\ensuremath{\mathrm{^{56}Co}}}
\newcommand{\Fe}{\ensuremath{\mathrm{^{56}Fe}}}
\newcommand{\Eej}{\ensuremath{E_{\mathrm{ej}}}}
\newcommand{\Mej}{\ensuremath{M_{\mathrm{ej}}}}
\newcommand{\Mst}{\ensuremath{M_{\mathrm{st}}}}
\newcommand{\Mni}{\ensuremath{M_{\Ni}}}
\newcommand{\Ha}{\ensuremath{{\mathrm{H}\alpha}}}
\newcommand{\kmps}{\ensuremath{\mathrm{km~s^{-1}}}}

\usepackage{graphicx}
\usepackage{txfonts}
\begin{document} 

   \authorrunning{Moriya et al.}
   \titlerunning{iPTF13ehe from a binary system}

   \title{
   Revealing the binary origin of Type~Ic superluminous supernovae
   through nebular hydrogen emission
   }

   \subtitle{}

   \author{Takashi J. Moriya\inst{1}
           \and
           Zheng-Wei Liu\inst{1}
           \and
           Jonathan Mackey\inst{2}
           \and 
           Ting-Wan Chen\inst{1}
           \and 
           Norbert Langer\inst{1}
          }

   \institute{Argelander Institute for Astronomy, University of Bonn,
              Auf dem H\"ugel 71, 53121 Bonn, Germany\\
              \email{moriyatk@astro.uni-bonn.de}
         \and 
  I.\ Physikalisches Institut, Universit\"at zu K\"oln, Z\"ulpicher Stra\ss{}e 77, 50937 K\"oln, Germany
             }

   \date{Received 06 October 2015 / Accepted 26 October 2015}

 
  \abstract{
  We propose that nebular \Ha\ emission as detected in the Type~Ic superluminous supernova iPTF13ehe stems from matter which is stripped from a companion star when the supernova ejecta collide with it. The temporal evolution, the line broadening, and the overall blueshift of the emission are consistent with this interpretation. We scale the nebular \Ha\ luminosity predicted for Type~Ia supernovae in single-degenerate systems to derive the stripped mass required to explain the \Ha\ luminosity of iPTF13ehe. We find a stripped mass of $0.1-0.9$ solar masses, assuming that the supernova luminosity is powered by radioactivity or magnetar spin down. Because a central heating source is required to excite the \Ha\ emission, an interaction-powered model is not favored for iPTF13ehe if the \Ha\ emission is from stripped matter. We derive a companion mass of more than 20 solar masses and a binary separation of less than about 20 companion radii based on the stripping efficiency during the collision, indicating that the supernova progenitor and the companion formed a massive close binary system. If Type~Ic superluminous supernovae generally occur in massive close binary systems, the early brightening observed previously in several Type~Ic superluminous supernovae may also be due to the collision with a close companion. Observations of nebular hydrogen emission in future Type~Ic superluminous supernovae will enable us to test this interpretation. 
  }


   \maketitle
%

\section{Introduction}
Superluminous supernovae (SLSNe) are a newly discovered class of supernovae (SNe)
(see \citealt[][]{gal-yam2012} for a review), and their origin is still a mystery.
In particular, the origin of Type~Ic SLSNe which do not show hydrogen features
in the early phases  \citep[e.g.,][]{quimby2011} has been actively discussed.
Several scenarios to make Type~Ic SLSNe bright are proposed,
e.g., magnetar spin-down \citep[e.g.,][]{kasen2010,dessart2012,inserra2013,chen2015,metzger2015},
large production of \Ni\ \citep[e.g.,][]{gal-yam2009,moriya2010,kasen2011,dessart2013,kozyreva2014,whalen2014},
and the interaction between SN ejecta and hydrogen-free dense circumstellar media
\citep[e.g.,][]{chevalier2011,moriya2012,ginzburg2012,chatzopoulos2012b,baklanov2015,sorokina2015}.
Revealing the progenitors and the power source of Type~Ic SLSNe is essential for a better understanding
of massive star evolution \citep{langer2012},
as well as for the possible use of Type~Ic SLSNe as a standardizable candle \citep{quimby2013,inserra2014}. 

Type~Ic SLSNe were thought to be hydrogen-free. However,
\citet[][Y15 hereafter]{yan2015} recently report the detection of nebular \Ha\ emission
at about 250~days after the $r$-band light-curve (LC) peak in the slowly declining Type Ic SLSN iPTF13ehe.
The LC evolution and spectral properties of iPTF13ehe are similar to those of SN~2007bi \citep[e.g.,][]{gal-yam2009},
which is suggested to be a pair-instability SN \citep[e.g.,][]{heger2002}.
Y15 estimate that about 10\% of Type~Ic SLSNe may have nebular \Ha\ emission as is observed
in iPTF13ehe.

Y15 suggest that the nebular \Ha\ emission in iPTF13ehe results from the interaction between
the SN ejecta and a hydrogen-rich circumstellar shell.
If the shell is located far from the progenitor,
it takes time for the SN ejecta to
reach the hydrogen-rich shell. Thus, the \Ha\ emission 
is observed in late phases. Y15 propose that the hydrogen-rich shell is formed
by the pulsational pair instability \citep[e.g.,][]{woosley2007}
of the progenitor and that the subsequent explosion may have resulted in iPTF13ehe.
The confinement of stellar wind by external photoionization may also result in the formation of a massive
circumstellar shell \citep{mackey2014}.
\citet{wang2015} invoke the necessity to simultaneously consider three suggested luminosity sources
in iPTF13ehe, i.e., \Ni, magnetars, and interaction.

In this Letter, we present an alternative view to interpret the nebular \Ha\ emission.
We suggest that the hydrogen observed in iPTF13ehe originates from the matter stripped from the
companion star contaminating the inner low-velocity layers of the SN ejecta.
The photosphere is in the outer high-velocity layers soon after explosion and we do not observe
emission from the inner layers in the early phase.
In the nebular phase when the inner layers are transparent,
emission from the contaminated inner layers becomes observable.
It has been argued that nebular \Ha\ emission from the stripped matter should be detectable
in Type~Ia SNe that originate from single-degenerate progenitor systems,
although this has not yet been observed
\citep[][]{mattila2005,leonard2007,shappee2013,lundqvist2013,lundqvist2015,maeda2014}.
The stripped mass in Type~Ia SNe is estimated to be less than $\sim 10^{-3}~\Msun$.
We apply the same idea to the case of iPTF13ehe and argue that
iPTF13ehe may be the first SN for which the emission of stripped matter from a companion star is detected.

\section{Stripped mass from the companion}
In this section, we estimate the mass required to be stripped from the companion star during the collision
in order to explain the hydrogen emission observed in iPTF13ehe.

\subsection{Scaling relation}
\citet{mattila2005} perform nebular spectral modeling of Type~Ia SNe
assuming that the inner layers of the ejecta of the W7 model \citep{nomoto1984}
are contaminated by solar-metallicity matter stripped from the companion.
They predict that the nebular \Ha\ luminosity from the stripped matter
is $\simeq 5\times 10^{36}~\mathrm{erg~s^{-1}}$ at 380~days after the explosion
for a stripped mass of $\Mst = 0.05~\Msun$.
We scale this result to estimate the stripped mass in iPTF13ehe.

We assume that the \Ha\ emission is mainly due to the non-thermal excitation of hydrogen
due to $\gamma$-rays from the nuclear decay of $\Co\rightarrow\Fe$
in the nebular phase \citep[e.g.,][]{mattila2005}.
The inner layers of the W7 model at 380~days are optically thin to
$\gamma$-rays. Thus, the \Ha\ luminosity ($L_\Ha$) is proportional to
$\tau_{\ej, \gamma}L_\gamma$, where $\tau_{\ej,\gamma}$ is the
$\gamma$-ray optical depth in the ejecta and $L_\gamma$ is the
$\gamma$-ray luminosity \citep[cf.][]{kozma1992}.
The $\gamma$-ray optical depth $\tau_{\ej, \gamma}$ is proportional to
$\rho_\ej R_\ej$ where $\rho_\ej$ is the ejecta density and
$R_\mathrm{ej}=(2\Eej/\Mej)^{1/2}t$ is the characteristic length scale in the ejecta.
Because the inner layers of SN ejecta have an almost-constant density profile \citep[e.g.,][]{kasen2010b},
we assume $\rho_\mathrm{ej}\propto \Mej^{5/2}\Eej^{-3/2}t^{-3}$.
Then, we obtain
$\tau_{\ej,\gamma}\propto \rho_\ej R_\mathrm{ej} \propto \Mej^{2}\Eej^{-1} t^{-2}$.
The late-phase $\gamma$-ray luminosity is $L_\gamma\propto \Mni \exp (-t/\tau_\Co)$ where $\Mni$ is the
initial \Ni\ mass and $\tau_\Co=111$~days is the \Co\ decay timescale.
Finally, the \Ha\ luminosity is proportional to the stripped mass \Mst.
Thus, we expect that the \Ha\ luminosity scales as
$L_\Ha\propto \Mst \Mej^{2}\Eej^{-1}\Mni t^{-2}\exp (-t/\tau_\Co)$.

Using the W7 parameters ($\Mej=1.4~\Msun$, $\Eej=1.3~\mathrm{B}$, and $\Mni=0.6~\Msun$)
and $L_\Ha\simeq 5\times 10^{36}~\mathrm{erg~s^{-1}}$ at $t=380~\mathrm{days}$
with $\Mst=0.05~\Msun$ \citep{mattila2005}, we obtain the scaling relation
\begin{eqnarray}
L_{\Ha}&\simeq&5\times 10^{44}
\left(\frac{\Mst}{\Msun}\right)
\left(\frac{\Mej}{\Msun}\right)^{2}
\left(\frac{\Eej}{\mathrm{B}}\right)^{-1} 
\left(\frac{\Mni}{\Msun}\right)
\nonumber
\\ 
&&
\ \ \ \ \ \ \ \ \ \ \ \ \ \ \ \ \ \ \ \ \
\times
\left(\frac{t}{\mathrm{day}}\right)^{-2}
\exp\left(-\frac{t}{111~\mathrm{days}}\right)
~\mathrm{erg~s^{-1}}.
\label{ha}
\end{eqnarray}
We apply this formula to iPTF13ehe in the next section.
Its validity and uncertainties for this application are discussed in Section~\ref{sec:valid}.

\subsection{The stripped mass in iPTF13ehe}
iPTF13ehe is observed to have $L_\Ha\simeq2\times10^{41}~\mathrm{erg~s^{-1}}$
at $\sim 250$~days after the LC peak (Y15).
The rise time is uncertain, so we adopt $\sim 110$~days based on the polynomial fit to the LC (Y15).
Given the redshift of iPTF13ehe (0.3434, Y15), we estimate that
the above \Ha\ luminosity is observed at about 270~days after the explosion in the rest frame.
Y15 find that the \Ha\ luminosity decreases by roughly 20\% from 270~days to 290~days
in the rest frame. Our interpretation that the stripped matter provides the \Ha\ luminosity
predicts about 28\% luminosity reduction during this period (Eq.~\ref{ha}).
This is consistent with the observation.

If iPTF13ehe is powered by \Ni, the required \Ni\ mass to obtain the peak luminosity is
about 15~\Msun\ (Y15). If we assume that iPTF13ehe is a pair-instability SN,
the expected ejecta mass and explosion energy are roughly 110~\Msun\ and 60~B,
respectively \citep[e.g.,][]{heger2002}. 
Substituting these properties for Eq.~(\ref{ha}), we obtain $\Mst\simeq 0.1~\Msun$.
If the large \Ni\ production is due to a core-collapse SN \citep[e.g.,][]{umeda2008},
the ejecta mass can be smaller than what we expect from the pair-instability SN \citep[e.g.,][]{moriya2010}.
A large explosion energy is still required to have the large \Ni\ production.
If we adopt $\Mej=70~\Msun$ and $\Eej=60~\mathrm{B}$, we obtain $\Mst=0.3~\Msun$. 

Even if iPTF13ehe is powered by a magnetar, a $\gamma$-ray deposition from the magnetar similar to that 
in the \Ni-powered model
is required to explain its slowly-declining LC. Thus, we estimate the stripped mass by
assuming $\Mni=15~\Msun$ in Eq.~(\ref{ha}) even in the magnetar model.
The main difference between the magnetar-powered and \Ni-powered models is
in the SN ejecta mass and energy. If we take typical ejecta mass (5~\Msun) and energy (1~B)
estimated for magnetar-driven SLSNe \citep[e.g.,][]{inserra2013,nicholl2015}, we obtain $\Mst\simeq 0.9~\Msun$.
However, the magnetar model for iPTF13ehe by \citet{wang2015} requires
$\Mej=35~\Msun$ and $\Eej=40~\mathrm{B}$. In this case, 
the stripped mass is $\Mst\simeq 0.7~\Msun$.

If the interaction between the SN ejecta and a hydrogen-free dense circumstellar medium
is the major luminosity source of iPTF13ehe, the SN explosion itself can be normal.
For instance, the Type~Ic SLSN 2010gx had
a spectrum that is similar to those observed in broad-line Type~Ic SNe \citep{pastorello2010}.
If we adopt typical broad-line Type~Ic SN properties
($\Eej=10~\mathrm{B}$, $\Mej=5~\Msun$, and $M_{\Ni}=1~\Msun$, e.g., \citealt{taddia2015}),
we obtain $\Mst\simeq 140~\Msun$ which is unrealistically high.
If we assume a higher ejecta mass, for example $\Mej=20~\Msun$, we obtain a lower stripped mass, $8~\Msun$.
A significant central heating source is necessary to excite hydrogen
to explain the nebular \Ha\ emission from the stripped mass.
We note that a reverse shock created by the interaction, which could propagate into the inner layers to excite
hydrogen, does not move rapidly inwards because the interaction forms a radiative, dense, cool shell \citep[e.g.,][]{sorokina2015}. 
Table~\ref{table:mass} summarizes the stripped masses estimated in this section.

\begin{table}
\caption{Estimated stripped mass \Mst\ for different progenitor models.
Values in parentheses for \Mni\ denote an energy input consistent with that provided by this mass of \Ni, even though this energy may be injected by a central engine.
} 
\label{table:mass}      
\centering                          
\begin{tabular}{lcccc}        
\hline\hline                 
model &\Mej/\Msun & \Eej/B & \Mni/\Msun   &  \Mst/\Msun \\
\hline         
            PISN &110 & 60 & 15 & 0.1     \\
            core-collapse~SN &70 & 60 & 15 & 0.3     \\
            magnetar&5 & 1 & (15) & 0.9     \\
            \citet{wang2015} &35 & 40 & (15) & 0.7    \\
            interaction&5 & 10 & 1 & 140     \\
\hline
            Type Ia (ref. model)   & 1.4 & 1.3 & 0.6 & 0.05 \\
\hline                                   
\end{tabular}
\end{table}

\subsection{Consistency check and uncertainties}\label{sec:valid}
In Eq.~(\ref{ha}), it is assumed that the inner layers of the SN ejecta are optically thin
to $\gamma$-rays as in the Type~Ia SN model
of \citet{mattila2005} at $t=380$~days.
Assuming a $\gamma$-ray opacity of $0.03~\mathrm{cm^2~g^{-1}}$ \citep{axelrod1980}, the
Type~Ia SN model has $\tau_{\ej,\gamma}\sim 10^{-2}$.
For the models at 270~days in the previous section,
we obtain $\tau_{\ej,\gamma}\sim 1$. Thus, the scaling relation
is still able to provide a rough estimate for the stripped mass.
Because the optical depth is around unity, the scaling relation 
may slightly underestimate the stripped mass.

Y15 estimate a \Ni\ mass of 2.5~\Msun\ in iPTF13ehe based on their nebular spectral modeling.
This \Ni\ mass is too small to explain the peak luminosity of iPTF13ehe, leading to
the suggestion that two or more luminosity sources power iPTF13ehe (Y15 and \citealt{wang2015}).
However, this \Ni\ mass estimate assumes that \Ni\ is irrelevant to the \Ha\ emission.
If the \Ha\ emission is powered by radioactive decay, the mass estimate of \Ni\ increases.
The estimated \Ni\ mass is also uncertain because of the uncertainties in the spectral modeling itself (Y15).
Thus, the \Ni-powered model with 15~\Msun\ of \Ni\ is not ruled out 
if the \Ha\ emission is emitted by the stripped matter.

\section{Constraint on the progenitor binary system}
\citet{liu2015} have recently studied the collision between SN ejecta and its companion star
in core-collapse SNe with different separations, companion masses, ejecta masses, and explosion energies
\citep[see also][]{hirai2014}.
They investigate how the stripping fraction $f\equiv \Mst/M_2$
(where $M_2$ is the companion mass before the collision) is affected by these parameters.
\citet{liu2015} only investigate the collision with the companion stars of 0.9 and 3.5~\Msun.
To constrain the massive progenitor system of iPTF13ehe, we perform numerical simulations
of the collision between the SN ejecta (110~\Msun\ and 60~B) and a massive main-sequence star
(50~\Msun\ with 10~\Rsun) by adopting the same method as in \citet{liu2015}.
The three-dimensional smoothed-particle hydrodynamics code \texttt{Stellar GADGET} \citep{pakmor2012} is used for the simulations.
We run simulations for two separations, 5~$R_2$ and 10~$R_2$, where $R_2$ is the companion radius.
The Roche-lobe radii for the mass ratios of 0.5 and 1 are 3.1~$R_2$ and 2.6~$R_2$, respectively
\citep{eggleton1983}.

The numerical simulations reveal stripped masses of
0.3~\Msun\ ($f\simeq 6\times 10^{-3}$) and 0.05~\Msun\ ($f\simeq 10^{-3}$) for the separations
of 5~$R_2$ and 10~$R_2$, respectively.
The stripped mass therefore reaches more than $0.1~\Msun$, as required from our estimate, and
the binary system with the 50~\Msun\ main-sequence companion is consistent with 
the \Ha\ luminosity observed in iPTF13ehe.
Assuming $f=(6-1)\times 10^{-3}$, the companion mass needs to be around $20-100~\Msun$
if $\Mst=0.1~\Msun$. Because the ejecta mass does not strongly affect the stripping fraction \citep{liu2015},
the companion mass is expected to be $50-300~\Msun$ if the ejecta mass is 70~\Msun\ ($\Mst=0.3~\Msun$).

The stripping fraction is proportional to \Eej\ \citep{liu2015}.
The magnetar model in \citet{wang2015} has a high explosion energy (40~B) and
the companion mass needs to be more than about 80~\Msun\ to have $\Mst\simeq 0.7~\Msun$.
If $\Eej\simeq 1~\mathrm{B}$ as estimated from the typical magnetar model,
the stripping fraction is likely to be $f\sim 10^{-4}$ or less.
Then, the companion mass needs to be larger than $\sim 10^4~\Msun$ ($\Mst=0.9~\Msun$),
which is extremely high.

The stripped mass strongly depends on the separation.
As the separation becomes larger, the angular size of the companion star from the SN ejecta becomes
smaller. Thus, less energy is ejected in the direction of the companion
and the mass stripping becomes less efficient.
The above stripping fraction is estimated for separations between 5 and $10~R_2$.
As is indicated in \citet{liu2015},
the stripping fraction drops significantly as the separation doubles
\citep[see also][]{hirai2014}.
If the separation is more than about $20~R_2$, the required companion mass is likely to become
more than $10^3$~\Msun\ in most of the models.

\section{Discussion}
\subsection{\Ha\ emission features}
So far, we have focused on the observed \Ha\ luminosity of iPTF13ehe.
Here we discuss other \Ha\ features of iPTF13ehe in the context of our stripping scenario.

\begin{figure}
  \centering
   \includegraphics[width=0.7\columnwidth]{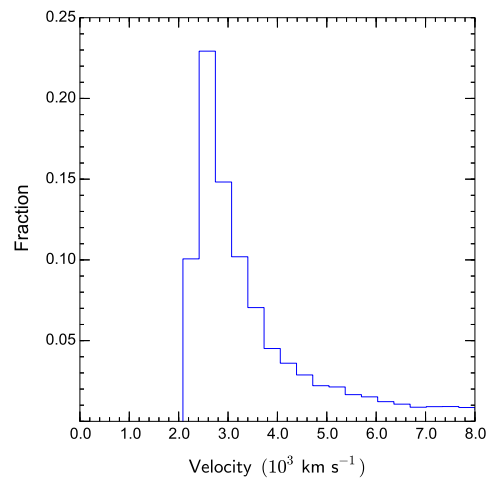}
   \caption{
   Final velocity distribution of the stripped matter from the main-sequence companion star in our numerical
   simulation with a separation of $5~R_2$.
   }
   \label{fig:vel}%
   \end{figure}

The \Ha\ emission in iPTF13ehe is observed to have a line broadening of $\simeq 4000~\kmps$
and an overall line shift of several $100~\kmps$ (Y15).
Because the stripped matter remains within the inner layers of the SN ejecta,
it is expected to emit at relatively low velocities ($\lesssim 1000~\kmps$ in the case of Type~Ia SNe,
e.g., \citealt{liu2012}). If the explosion energy is higher, stripped matter can end up with
higher velocities and we expect to observe broader emission lines, as seen in iPTF13ehe.
In Fig.~\ref{fig:vel}, we show the velocity distribution of the stripped matter
in our impact simulation with a separation $5~R_2$.
The typical velocity of the stripped matter is around 3000~\kmps,
while the typical velocity of the ejecta is around 8000~\kmps.
The velocities of the stripped matter are actually slower than the typical ejecta
velocity and they are close to the line broadening observed in iPTF13ehe.

If the companion star is located between the progenitor and the observer at the time of the explosion,
the collision occurs in the ejecta moving toward the observer. Thus, we expect
that the stripped matter has an overall velocity toward the observer and
an observer measures blueshifted \Ha\ emission.
If the companion is in the opposite direction,
the emission is likely to have an overall redshift. Thus, for emission from
the stripped matter, we expect to eventually observe an overall redshift in other Type~Ic SLSNe.
Detailed modeling of the nebular spectra based on impact simulations similar to \citet{liu2015}
is required to further investigate the expected emission properties.

\subsection{Early brightening observed in Type~Ic SLSNe}
Some Type~Ic SLSNe are known to have an early brightening before the main LC rise
\citep{nicholl2015b,leloudas2012}.
This early brightening has been related to, e.g.,
circumstellar interaction \citep{moriya2012} and magnetar spin-down \citep{kasen2015}.
However, if Type~Ic SLSNe originate from binary systems, this early emission may be related
to the collision between the SN ejecta and the companion \citep{kasen2010b}.
We expect an early luminosity of $\sim (1-5) \times 10^{43}~\mathrm{erg~s^{-1}}$
within 10~days after the explosion from the analytic formula of \citet{kasen2010b},
assuming $\Mej=110~\Msun$, an ejecta velocity of $13000~\kmps$ (Y15), and a separation of $10^{13}$~cm.
This is consistent with the early luminosity observed in Type~Ic SLSNe \citep{leloudas2012,nicholl2015b}.
The luminosity estimate is not sensitive to \Mej\ and we expect similar
luminosity in the magnetar model \citep{kasen2010b}.
Because the general likelihood to observe an early brightening by the companion star in Type~Ic SNe is estimated
to be low \citep{moriya2015}, the frequent observations of an early brightening in
Type~Ic SLSNe may also imply that they preferentially occur in close binary systems.
Early brightening by impact onto a companion star strongly depends on the direction to the observer, while
\Ha\ emission from stripped matter can be observed from all directions \citep[e.g.,][]{kasen2010b,liu2015b}.
Thus, \Ha\ emission may be observed more frequently. For example, if we assume a companion viewing angle of
about 10 degrees as is nearly the case for $5~R_2$, about 10\% of randomly oriented observers are in a preferred 
direction to find the early brightening \citep{kasen2010b}.

\subsection{The progenitor of iPTF13ehe and its binary evolution}
Binary evolution has been suggested to be important for SLSN progenitors \citep[e.g.,][]{justham2014}.
The fact that we find a possible signature of the companion star in Type~Ic SLSN iPTF13ehe suggests
that binary evolution can indeed play a key role in their progenitors.
For example, binarity can help progenitors to have a large angular momentum through tidal interactions
\citep[e.g.,][]{demink2009,song2015} and mass transfer \citep[e.g.,][]{demink2013}.
Rotationally-induced mixing may reduce the initial mass needed to obtain PISNe
\citep[e.g.,][]{yoon2012,chatzopoulos2012,yusof2013}.
Rapid rotation is also essential in the magnetar-powered model as its energy source is rotation.

Y15 estimate that about 10\% of Type~Ic SLSNe may have nebular \Ha\ emission.
This may indicate that insufficient mass to have detectable nebular \Ha\ emission
is stripped from the companion in about 90\% of Type~Ic SLSNe currently observed.
For example, if the typical separation is larger than about $20~R_2$,
we do not expect sufficient stripping to observe the emission, as discussed.
Alternatively, many progenitors may be effectively single stars.
This does not mean that they are single stars from the beginning.
A stellar merger leads to an apparent single star \citep[cf.][]{justham2014} and
to the increase of its rotational velocity \citep[cf.][]{demink2013}.

\section{Conclusions}
We suggest that the \Ha\ emission observed during the nebular phases of Type~Ic SLSN iPTF13ehe originates
from matter which is stripped off a hydrogen-rich companion star of the SN progenitor.
Matter near the surface of the companion star is stripped when the SN ejecta collide
with the companion and the hydrogen-rich stripped matter contaminates
the inner low-velocity layers of the SN ejecta.
Emission from the contaminated ejecta can only be observed during the nebular phases when the inner layers
are transparent.
The temporal evolution, the line broadening, and the overall blueshift of the observed \Ha\ emission
in iPTF13ehe are consistent with emission from stripped matter.
By scaling the predicted \Ha\ luminosity based on the W7 model of Type~Ia SNe,
we estimate a stripped mass in iPTF13ehe between 0.1 and 0.9~\Msun\ (Table~\ref{table:mass}).
We do not expect nebular \Ha\ emission from the stripped matter
if iPTF13ehe is powered by the interaction between the SN ejecta and a dense circumstellar medium
because a central heating source is required to excite hydrogen located at the inner low-velocity layers
of the SN ejecta.

We perform numerical simulations of the mass stripping using the same technique as in \citet{liu2015}.
We assume $\Mej=110~\Msun$ and $\Eej=60~\mathrm{B}$ and put a main-sequence companion star
with 50~\Msun\ and 10~\Rsun\ at a separation of $5~R_2$ and $10~R_2$.
We find stripped masses of 0.3~\Msun\ ($5~R_2$) and 0.05~\Msun\ ($10~R_2$).
Our mass stripping efficiency implies that the companion mass is larger than about 20~\Msun\ (\Ni-powered models)
or more than about 80~\Msun\ (magnetar spin-down models).
Our results imply that the progenitor evolved in a massive binary system and that stellar binarity may 
play a critical role in the evolution of Type~Ic SLSN progenitors.
If Type~Ic SLSN progenitors are typically in close binary systems, the early brightening
sometimes observed in them may also be related to the collision.

If our interpretation of the origin of the \Ha\ emission is correct,
we expect some diversity in the \Ha\ emission feature. Especially, the overall line shift,
which is a blue shift in the case of iPTF13ehe, is likely affected by the viewing angle.
If the companion star is located behind the SN progenitor at the time of explosion,
an overall redshift of the \Ha\ emission may be observed.
It will be also interesting to obtain the \Ha\ detection rate separately
in slowly-declining and fast-declining Type~Ic SLSNe in future SLSN studies.
A large number of observations of late-phase emission from Type~Ic SLSNe can test our interpretation and
reveal the origin and nature of mysterious Type~Ic SLSNe.

\begin{acknowledgements}
TJM is supported by Japan Society for the Promotion of Science Postdoctoral Fellowships for
Research Abroad (26 \textperiodcentered 51).
JM acknowledges support from the Deutsche Forschungsgemeinschaft priority program 1573, Physics of the Interstellar Medium. 
\end{acknowledgements}


\end{document}